\documentclass [12pt]{article}
\usepackage {graphicx}
\usepackage[american]{babel}

\begin{document}

\title{Synchronization of forced reactively coupled van der Pol oscilators }

\author{A.P. Kuznetsov$^{1}$, L.V. Turukina$^{1}$, \\ N.Yu. Chernyshov$^{2}$ and Yu.V. Sedova$^{1}$}

\maketitle
\begin{center}
\textit{$^{1}$ Kotel'nikov's Institute of Radio-Engineering and Electronics of RAS, Saratov Branch,\\
Zelenaya 38, Saratov, 410019, Russian Federation\\
$^{2}$ Saratov State University, \\
Astrachanskaya 83, Saratov, 410012, Russian Federation}
\end{center}

\begin{abstract}
Synchronization of forced reactively coupled van der Pol oscillators is investigated in the phase approximation. We discuss essential features of the reactive coupling. Bifurcation mechanisms for the destruction of complete synchronization and possible quasi-periodic regimes of different types are revealed. Regimes when autonomous oscillators demonstrate frequency locking and beating regimes with incommensurate frequencies are considered and compared.
\end{abstract}


\section{Introduction}

Synchronization is a common phenomenon in physics, chemistry, biology and other fields of science and technology \cite{b1, b2}. Nowadays, different examples and variety of synchronization in ensembles of oscillators are topics of the research interest \cite{b3}. It has been found that systems containing even a small number of elements demonstrate a great variety of different types of dynamics \cite{b4, b5}.  However, a complete picture of synchronization is still not constructed. One of the fundamental problems is a problem of influence of an external signal to coupled oscillators.  Some new aspects of this problem have been recently revealed for dissipative coupling. A corresponding phase equation which generalizes classical Adler equation has been obtained in \cite{b6}. A case when two autonomous oscillators are locked has been investigated. A complete synchronization region for two oscillators with an external force has been detected, and degenerated saddle-node bifurcation of equilibrium states responsible for the destruction of such synchronization has been described. It has been shown that three-frequency quasi-periodicity appears in the phase model through a saddle-node bifurcation of invariant curves. External force synchronization has been investigated in \cite{b7} in situation when a locking regime in system of autonomous oscillators changes to a beating regime with incommensurable frequencies. An influence of asymmetry of coupling to synchronization picture has been investigated in \cite{b8}. Excitation of a chain of three phase oscillators with dissipative coupling has been considered in \cite{b9}. Synchronization mechanisms beyond the phase approximation have been discussed in \cite{b10}. A possibility of both the regimes observed in the phase model and a new kind of synchronization via suppression of oscillations has been shown.

These results, however, refer to the simplest case when the coupling is dissipative. There is another type of coupling, namely reactive (or conservative) coupling. This is a situation when the coupling is realized directly through values of the variables and not through their rates \cite{b1, b11}. Reactive coupling occurs in many interesting problems. An actual example of such coupling is ion traps \cite{b12}. In such traps ions are confined using variable microwave fields, which restrict a magnitude of radial oscillations of the ions and a constant electric field limiting the axial motions. Laser radiation of different frequencies provides instability and dissipation into the system. Reactive type of the coupling is caused by Coulomb repulsion between the ions. Moreover a reactive coupling arises in the array of coupled nanomechanical cavities as well as in some optical problems \cite{b13}.

From a theoretical point of view, the reactive coupling is an essentially more delicate phenomenon than the dissipative coupling \cite{b1, b11}. So in the simplest case of two coupled oscillators, equations in the phase approximation can be obtained only if we deal with second order of smallness in the coupling parameter. For this, in contrast to the dissipative coupling, it is necessary to take into account deviation of their orbits from the unperturbed states \cite{b11}. Phase bistability is possible for the reactive coupling when synchronization between oscillators may be in-phase as well as anti-phase. At the same time a problem of forced synchronization of reactively coupled oscillators has not yet been fully investigated. We should note the work \cite{b14} having a style of the computer experiment. This work has demonstrated a possibility of two- and three-frequency quasi-periodicity and chaos in such a system. In the more recent paper \cite{b15}, a structure of the parameter plane "frequency - amplitude of an external signal" has been investigated, but the most significant and complex region of small signal amplitudes has not been considered. At the same time, a phase model has not been constructed and the authors have investigated only the case when autonomous oscillators are locked. These works describe the variety of possible effects and bifurcations in such system not well enough. Therefore, we discuss in this paper the following questions:
\begin{itemize}
  \item How to write the correct phase equations for the forced system in case of the reactive coupling between oscillators?
  \item What is the structure of complete synchronization region for such forced oscillators, and what bifurcation is responsible for the destruction of this region?
  \item What effects do occur in the presence of phase bistability in case of an external forcing?
  \item How are the various two-frequency and three-frequency regimes embedded in the parameter space?
\end{itemize}

An important moment is a parallel consideration of two possible situations. The first one is when autonomous oscillators are locked, and the second one is when they demonstrate oscillations in the beating regime.

\section{Phase equations of reactively coupled oscillators}

Consider a system of two reactively coupled van der Pol oscillators excited by an external harmonic signal:
\begin{equation}
\label{eq1}
\setlength\arraycolsep{2pt}
\begin{array}{l}
  \ddot{x} - (\lambda - x^{2}) \dot{x} + (1- \displaystyle \frac{\Delta}{2}) x + \varepsilon (x - y) = B \sin \omega t, \\
  \\
  \ddot{y} - (\lambda - y^{2}) \dot{y} + (1 + \displaystyle \frac{\Delta}{2}) y + \varepsilon (y - x)=0.
\end{array}
\end{equation}
Here $\lambda$  is the control parameter,  $\Delta$  is the frequency detuning between oscillators,  $\varepsilon$  is the parameter of reactive coupling, $B$ is the signal amplitude, and $\omega$  is the signal frequency. Central frequency of the oscillators is taken as unity.

We assume that control parameter, coupling value and signal frequency detuning from the unit frequency are small. In this case we can use the method of slowly varying amplitudes \cite{b1}. We represent the dynamic variables as
\begin{equation}
\label{eq2}
\setlength\arraycolsep{2pt}
\begin{array}{c}
  x = a(t) e^{i \omega t} + a^{\ast} (t) e^{- i \omega t}, \quad
  y = b(t) e^{i \omega t} + b^{\ast} (t) e^{- i \omega t}.
\end{array}
\end{equation}
and apply standard additional conditions:
\begin{equation}
\label{eq3}
 \setlength\arraycolsep{2pt}
\begin{array}{c}
 \dot{a}(t) e^{i \omega t} + \dot{a}^{\ast} (t) e^{- i \omega t}=0, \quad
 \dot{b}(t) e^{i \omega t} + \dot{b}^{\ast} (t) e^{- i \omega t}=0.
\end{array}
\end{equation}
where $a(t)$  and $b(t)$  are the complex amplitudes. Let us substitute the relations (\ref{eq2}), (\ref{eq3}) in the equations (\ref{eq1}), multiply the obtained expressions by  $e^{-i \omega t}$ and average the result for eliminating rapidly oscillating terms. Also we use the relation $B \sin \omega t = \displaystyle  \frac{B(e^{i \omega t} - B e^{-i \omega t})}{2}$. After some mathematical transformations we obtain the Landau-Stuart equations:
\begin{equation}
\label{eq4}
\setlength\arraycolsep{2pt}
\begin{array}{l}
  2\dot{a}=\lambda a + i \displaystyle \frac{1 - \omega^{2} - \displaystyle \frac{\Delta}{2}}{\omega} a - |a|^{2} a + i \varepsilon (a - b) - \displaystyle \frac{B}{2 \omega}, \\
  \\
  2\dot{b}=\lambda b - i \displaystyle \frac{1 - \omega^{2} + \displaystyle \frac{\Delta}{2}}{\omega} b - |b|^{2} b + i \varepsilon (b - a).
\end{array}
\end{equation}
Now we introduce the frequency detuning of the signal $\Omega$  from the central frequency: $\omega = 1 + \Omega$. Since $\omega \approx 1$, we can set $\displaystyle \frac{1 - \omega^{2} \mp \displaystyle \frac{\Delta}{2}}{\omega} \approx - 2\Omega \mp \displaystyle \frac{\Delta}{2}$.  Furthermore, the parameter  $\lambda$ in Eqs.(\ref{eq4}) can be eliminated by renormalization: $t \rightarrow \displaystyle \frac{\tau}{\lambda}$, $\Omega \rightarrow \displaystyle \frac{\Omega}{\lambda}$, $\Delta \rightarrow \displaystyle \frac{\Delta}{\lambda}$, $\varepsilon \rightarrow \displaystyle \frac{\varepsilon}{\lambda}$ and $B \rightarrow \displaystyle \frac{B}{\lambda}$. As a result, we obtain:
\begin{equation}
\label{eq5}
\setlength\arraycolsep{2pt}
\begin{array}{l}
  2\dot{a}= a + i (2\Omega + \displaystyle \frac{\Delta}{2}) a - |a|^{2} a + i \varepsilon (a - b) - \displaystyle \frac{B}{2}, \\
  \\
  2\dot{b}= b - i (2\Omega - \displaystyle \frac{\Delta}{2}) b - |b|^{2} b + i \varepsilon (b - a).
\end{array}
\end{equation}
Let us introduce the real amplitudes $r_{1,2}$  and phases of the oscillators with respect to the external signal $\psi_{1,2}$  using the relations $a = r_{1} e^{i\psi_{1}}$ and  $b = r_{2} e^{i\psi_{2}}$. Then Eqs.(\ref{eq5}) yield:
\begin{equation}
\label{eq6}
\setlength\arraycolsep{2pt}
\begin{array}{l}
 2\dot{r}_{1} = r_{1} - r_{1}^{3} - \varepsilon r_{2} \sin \theta - \displaystyle \frac{B}{2} \cos \psi_{1},  \\
\\
2\dot{r}_{2} = r_{2} - r_{2}^{3} + \varepsilon r_{1} \sin \theta, \\
  \\
 2\dot{\psi}_{1} = -2\Omega - \displaystyle \frac{\Delta}{2}  + \varepsilon - \displaystyle \frac{r_{2}}{r_{1}} \varepsilon \cos \theta + \displaystyle \frac{B}{2} \sin \psi_{1},  \\
\\
2\dot{\psi}_{2} = -2\Omega + \displaystyle \frac{\Delta}{2} + \varepsilon - \displaystyle \frac{r_{1}}{r_{2}} \varepsilon \cos \theta.
\end{array}
\end{equation}
Here $\theta = \psi_{1} - \psi_{2}$  is the relative phase of the oscillators.

Let us now to take into account the deviation of orbits from stationary unperturbed values  $r_{1} = r_{2} = 1$. For this we set  $r_{1,2} = 1 + \tilde{r}_{1,2}$, where $\tilde{r}_{1,2}$ are the small perturbations, and rewrite  first and second equations in Eqs.(\ref{eq6}) as:
\begin{equation}
\label{eq7}
\setlength\arraycolsep{2pt}
\begin{array}{l}
 2\dot{r}_{1} = -2 \tilde{r}_{1} - \varepsilon \sin \theta - \displaystyle \frac{B}{2} \cos \psi_{1},  \quad 2\dot{r}_{2} = -2 \tilde{r}_{2} + \varepsilon \sin \theta.
 \end{array}
\end{equation}
In these equations the amplitude perturbations are strongly damped, so there is rapid convergence to the stationary states:
\begin{equation}
\label{eq8}
\setlength\arraycolsep{2pt}
\begin{array}{l}
 \tilde{r}_{1} = - \displaystyle \frac{\varepsilon}{2} \sin \theta - \displaystyle \frac{B}{2} \cos \psi_{1},  \quad \tilde{r}_{2} = \displaystyle \frac{\varepsilon}{2} \sin \theta.
 \end{array}
\end{equation}

Substituting these expressions in third and fourth equations in Eqs.(\ref{eq6}) and retaining only the first and second order terms of the coupling parameter $\varepsilon$ gives:
\begin{equation}
\label{eq9}
\setlength\arraycolsep{2pt}
\begin{array}{l}
  \dot{\psi}_{1} = - \displaystyle \frac{\Delta}{4}  - \Omega  + \displaystyle \frac{\varepsilon}{2}(1 - \cos \theta) - \displaystyle \frac{\varepsilon^{2}}{4} \sin 2\theta + \displaystyle \frac{B}{4} \sin \psi_{1},    \\
  \\
 \dot{\psi}_{2} =  \displaystyle \frac{\Delta}{4} - \Omega + \displaystyle \frac{\varepsilon}{2}(1 - \cos \theta) + \displaystyle \frac{\varepsilon^{2}}{4} \sin 2\theta.
\end{array}
\end{equation}
Here we assume that the amplitude $B$ is of order $\varepsilon^{2}$ and neglect the term $B \varepsilon$. Physical explanation is that the external signal perturbs only phases of the oscillators and practically does not perturb amplitudes. Subtracting Eqs.(\ref{eq9}) from each other, we obtain:
\begin{equation}
\label{eq10}
\setlength\arraycolsep{2pt}
\begin{array}{l}
  \dot{\theta} = - \displaystyle \frac{\Delta}{2}  - \displaystyle \frac{\varepsilon^{2}}{2} \sin 2\theta + \displaystyle \frac{B}{4} \sin \psi_{1},    \\
  \\
 \dot{\psi}_{1} = - \displaystyle \frac{\Delta}{4}  - \Omega  + \displaystyle \frac{\varepsilon}{2}(1 - \cos \theta) - \displaystyle \frac{\varepsilon^{2}}{4} \sin 2\theta + \displaystyle \frac{B}{4} \sin \psi_{1}.
\end{array}
\end{equation}

Eqs.(\ref{eq10}) are the desired phase equations for forced reactively coupled oscillators. In the absence of coupling $\varepsilon=0$ we obtain phase equation of excitation of the first oscillator [1]. On the other hand, in the absence of external signal  $B=0$  Eqs.(\ref{eq10}) can be written as the well-known equation for the relative phase $\theta = \psi_{1} - \psi_{2}$ of two reactively coupled oscillators \cite{b11}:
\begin{equation}
\label{eq11}
\setlength\arraycolsep{2pt}
\begin{array}{l}
  \dot{\theta} = - \displaystyle \frac{\Delta}{2}  - \displaystyle \frac{\varepsilon^{2}}{2} \sin 2\theta.
\end{array}
\end{equation}
This equation describes an ability of the beating regime at $|\Delta| > \varepsilon^{2}$  and mutual locking of oscillators at  $|\Delta| < \varepsilon^{2}$. We can see also from Eq.(\ref{eq11}) that two oscillators may exhibit both the in-phase ($\theta \approx 0$) and anti-phase ($\theta \approx \pi$) stable synchronization. There are also two unstable synchronous regimes in the intermediate values of $\theta$ \cite{b11}.

Model (\ref{eq10}) was obtained by a number of restrictions on the parameters. However, it turns out, as we will show, quite effective in a wider area.

\section{Analysis of equilibrium states. Saddle-node bifurcations}

One of the basic regimes in the forced system is a complete synchronization when an external signal is locking for both oscillators. In this case relative phases of the oscillators with respect to the external signal $\psi_{1}$  and $\psi_{2}$  are constant in time. This regime corresponds to a stable equilibrium state of the system (\ref{eq10}). Let us find the region of complete synchronization on the parameter plane and discuss the mechanisms of its destruction.

In the case of dissipative coupling, one stable node, one unstable node and two saddles coexist in the phase model \cite{b6}. Destruction of the complete synchronization is caused by the specific variant of saddle-node bifurcation when all the four equilibrium states merge in pairs simultaneously and disappear, and stable and unstable invariant curves occur from their manifolds \cite{b6}. In the case of reactive coupling, the mechanisms of synchronization and its destruction are changing and become more complicated.

To find the equilibrium states, we set  $\dot{\psi}_{1} = \dot{\psi}_{2} = 0$  in Eqs.(\ref{eq10}):
\begin{equation}
\label{eq12}
\setlength\arraycolsep{2pt}
\begin{array}{l}
   - \displaystyle \frac{\Delta}{2}   - \displaystyle \frac{1}{2} \varepsilon^{2} \sin 2\theta + b \sin \psi_{1} = 0,    \\
  \\
 - \Omega - \displaystyle  \frac{\Delta}{4} + \displaystyle \frac{\varepsilon}{2}(1 - \cos \theta) - \displaystyle \frac{1}{4} \varepsilon^{2} \sin 2\theta + b \sin \psi_{1} = 0.
\end{array}
\end{equation}
Here  $b =\displaystyle \frac{B}{4}$. The perturbed matrix of the system (\ref{eq10}) is
\begin{equation}
\label{eq13}
 \hat{M} = \left(
             \begin{array}{cc}
               - \varepsilon^{2} \cos 2\theta & b \cos \psi_{1} \\
               \displaystyle \frac{\varepsilon}{2} \sin \theta - \displaystyle \frac{\varepsilon^{2}}{2} \cos 2\theta & b \cos \psi_{1} \\
             \end{array}
           \right)
\end{equation}
The condition of saddle-node bifurcation is zero Jacobian of this matrix, i.e. $\det \hat{M} = \cos \psi_{1} [ \varepsilon \cos 2\theta + \sin \theta] =  0$. The obtained equation can be split into two equations:
\begin{equation}
\label{eq14}
\varepsilon \cos 2\theta + \sin \theta = 0,  \quad \cos \psi_{1} = 0.
\end{equation}
The first one leads to the relation
\begin{equation}
\label{eq15}
\sin \theta = \displaystyle \frac{1 \mp\sqrt{1 + 8 \varepsilon^{2}}}{4 \varepsilon}.
\end{equation}
At $\varepsilon < 1$  it has a solution only if we use an upper sign. Subtracting one equation from another in Eqs.(\ref{eq12}) gives the following conditions for the saddle-node bifurcation:
\begin{equation}
\label{eq16}
\setlength\arraycolsep{2pt}
\begin{array}{c}
   \Omega = \displaystyle \frac{\Delta}{4} + \displaystyle \frac{\varepsilon}{2} (1 - \cos \theta) + \displaystyle \frac{1}{4} \varepsilon^{2} \sin 2\theta,    \\
  \\
 \theta_{1} = \arcsin (\displaystyle \frac{1 - \sqrt{1 + 8 \varepsilon^{2}}}{4 \varepsilon}), \quad  \theta_{2} = \pi - \arcsin (\displaystyle \frac{1 - \sqrt{1 + 8 \varepsilon^{2}}}{4 \varepsilon}).
\end{array}
\end{equation}

Conditions (\ref{eq16}) define the locking band bounded by two vertical lines on the ($\Omega$, $b$)  parameter plane, so that $\Omega_{min} < \Omega < \Omega_{max}$  (see Fig.1a). Width of the locking band is determined by the coupling parameter. It is easy to show that at small coupling values it has a width approximately equal to  $\varepsilon$.
\begin{figure}[!ht]
\centerline{
\includegraphics[scale=0.45]{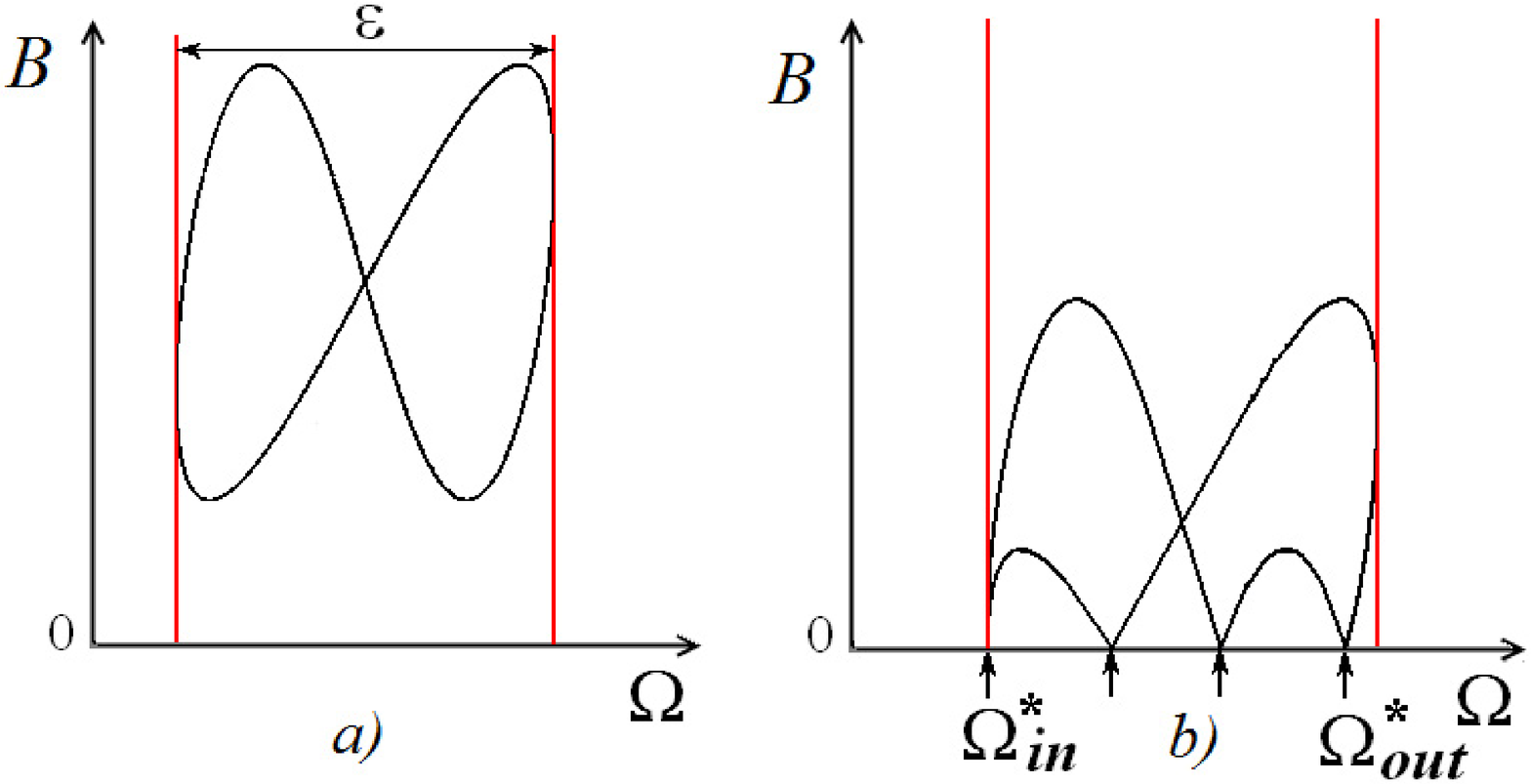}\\}
\caption{Lines (\ref{eq16}) and (\ref{eq17}) for $\varepsilon = 0.3$; a) beating regime of autonomous oscillators, $\Delta = 0.3$; b) locking regime of autonomous oscillators, $\Delta = 0.05$.  $\Omega^{*}_{in}$  and  $\Omega^{*}_{out}$  are the eigenfrequencies of the in-phase and out-phase modes. Unsigned arrows indicate locking frequencies in the unstable regimes.}
\end{figure}

Now consider the second condition  $\cos \psi_{1} = 0$. Then  $\sin \psi_{1} = \pm 1$  and Eqs.(\ref{eq12}), (\ref{eq13}) give
\begin{equation}
\label{eq17}
2b = \pm (\Delta + \varepsilon^{2} \sin 2\theta),  \quad \Omega = \displaystyle \frac{\Delta}{4} + \displaystyle \frac{\varepsilon}{2} (1 - \cos \theta) + \displaystyle \frac{1}{4} \varepsilon^{2} \sin 2\theta.
\end{equation}

The lines (\ref{eq17}) have a different configuration depending on whether the autonomous oscillators are in beating ($\Delta > \varepsilon^{2}$) or in locking ($\Delta < ^{2}$) regimes. Both of these cases are shown in Fig.1 on the left and right side, respectively. Here and below, we need to select the normalized values of the coupling parameter $\varepsilon$. We use the value $\varepsilon = 0.3$. The smallness of the parameter $\varepsilon^{2}$ (in this case $\varepsilon^{2} = 0.09$) is a condition of the effectiveness of the phase model. At the same time, the choice of $\varepsilon = 0.3$ will further demonstrate certain differences of the original system from the phase model.

In the case of beating regime  $\Delta > \varepsilon^{2}$  the condition $b \neq 0$  is always satisfied. Therefore, the line (\ref{eq17}) does not touch frequency axis and the complete synchronization region has an amplitude threshold in the beating regime of autonomous oscillators. This line forms a closed curve in the form of the "eight" number, Fig.1a. It is easy to show that the vertical lines (\ref{eq16}) touch this curve.

In the case of locking regime of autonomous oscillators the configuration of lines becomes more complicated, Fig.1b. Now they have four characteristic tips at  $b = 0$  marked by arrows in Fig.1b. If  the signal amplitude tends to zero  $b \rightarrow 0$,  Eqs.(\ref{eq17}) give the conditions for locking frequencies of two autonomous reactively coupled oscillators. The values marked in Fig.1b by arrows correspond to the eigenfrequencies of these modes.   $\Omega^{*}_{in}$ and $\Omega^{*}_{out}$  indicate the frequencies of stable in-phase and out-phase regimes, and similar points for unstable regimes are marked by unsigned arrows. Thus, the in-phase and out-phase mode are locked at different frequencies.

\section{Complete picture of bifurcations. Andronov-Hopf bifurcation}

However, it turns out that there are two different types of locking of both oscillators by an external signal in a system with reactive coupling. The first one is defined by a stable equilibrium of the system (\ref{eq10}) and has been discussed above. The corresponding phase portrait is presented in Fig.2, fragment 1. The second type of locking corresponds to a stable limit cycle of the system (\ref{eq10}) shown in Fig.2, fragment 6. In this locking regime, the phases of oscillators are not constant and oscillate near some equilibrium value. Therefore, even in the phase approximation, a new line defined by the period of phase oscillations appears in the spectrum of oscillators.

From this it follows that the Andronov-Hopf bifurcation (the stable limit cycle birth bifurcation) is possible in the phase system with the reactive coupling. Therefore, we consider a more complete bifurcation picture of the analyzed system shown in Fig.3 and revealed numerically.\footnote{Note that analytical investigation in Sec.2 is an essential supplement to numerical analysis due to the complexity of the bifurcation picture.} This figure indicates also two cases: beating and locking regimes of autonomous oscillators. For the beating regime (Fig.3a), there are two branches of saddle-node bifurcation lines $SN$ revealed in the analytical study. They touch each other at degenerate cusp points $DC$. The saddle-node bifurcation lines contain the segments $SN_{1}$  where merging of stable equilibrium and saddle occurs (solid line) and the segments  $SN_{2}$  where merging of unstable equilibrium and saddle takes place (dashed line). The Andronov-Hopf bifurcation $H$ is also possible. Supercritical bifurcation $H_{1}$  is responsible for emergence of a stable limit cycle, and subcritical bifurcation $H_{2}$  for  is responsible disappearance of an unstable limit cycle. The Andronov-Hopf bifurcation lines have common points with the saddle-node bifurcations line, the Bogdanov-Takens points $BT$. Grey color in Fig.3 shows the regions where the system has at least one equilibrium state.
\begin{figure}[!ht]
\centerline{
\includegraphics[scale=0.35]{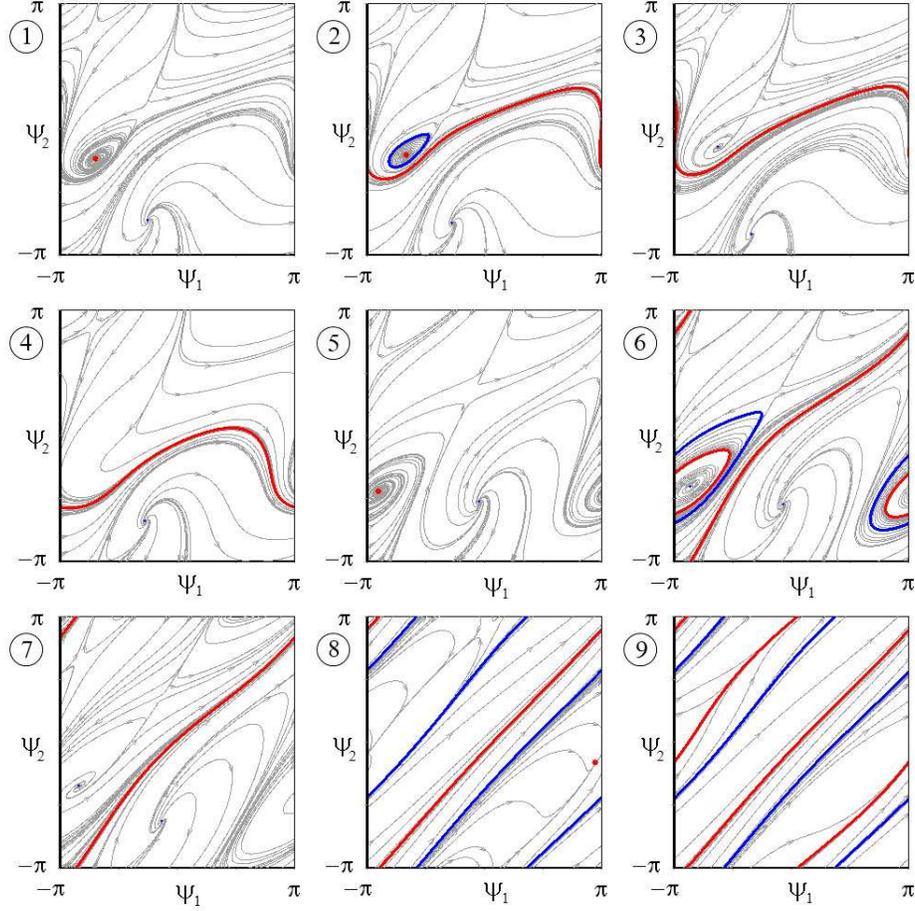}\\}
\caption{Phase portraits of the system (\ref{eq10}). Numbers correspond to the points in Fig.3. Values of the parameters are $\Delta = 0.3$, $\Omega = 0.21$ and $B = 0.8$ (Fragment 1); $\Delta = 0.3$, $\Omega = 0.21$ and $B = 0.75$ (Fragment 2); $\Delta = 0.3$, $\Omega = 0.22$ and $B = 0.67$ (Fragment 3); $\Delta = 0.3$, $\Omega = 0.17$ and $B = -0.66$ (Fragment 4); $\Delta = 0.05$, $\Omega = 0.15$ and $B = 0.5$ (Fragment 5); $\Delta = 0.05$, $\Omega = 0.15$ and $B = 0.355$ (Fragment 6); $\Delta = 0.05$, $\Omega = 0.16$ and $B = 0.23$ (Fragment 7); $\Delta = 0.05$, $\Omega = 0.29$ and $B = 0.7$ (Fragment 8); $\Delta = 0.05$, $\Omega = 0.31$ and $B = 0.035$ (Fragment 9).}
\end{figure}
\begin{figure}[!ht]
\centerline{
\includegraphics[scale=0.45]{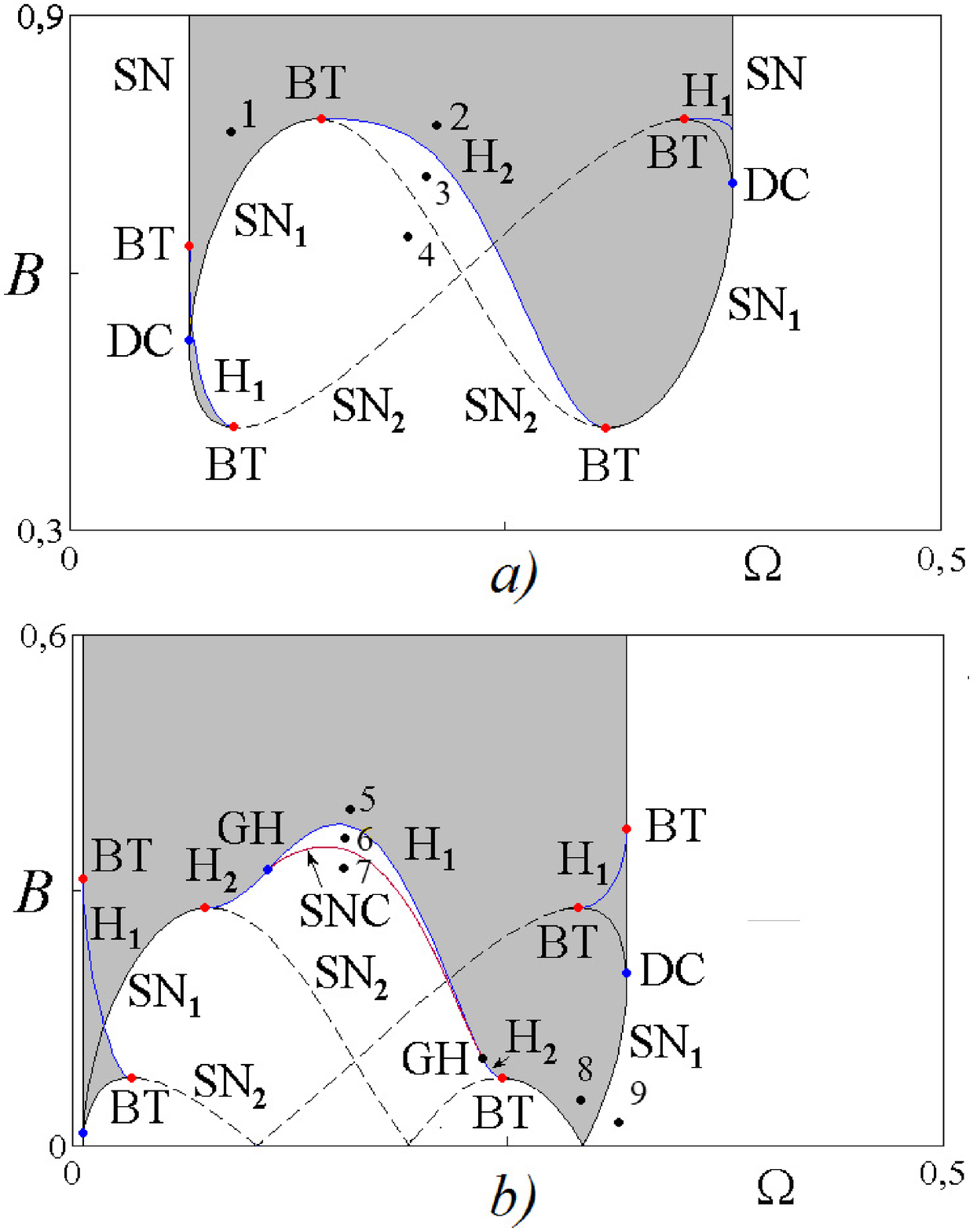}\\}
\caption{Bifurcation lines and points of the system (\ref{eq10}) on the ($\Omega$, $B$)  parameter plane for  $\varepsilon = 0.3$;     a) beating regime, $\Delta = 0.3$; b) locking of autonomous oscillators, $\Delta = 0.05$. $SN$ is the saddle-node bifurcation of equilibrium states, $H_{1}$ is  the supercritical Andronov-Hopf bifurcation, $H_{2}$ is the subcritical Andronov-Hopf bifurcation, $SNC$ is the saddle-node bifurcation of limit cycles, $DC$ is the degenerate cusp point, $BT$ is the Bogdanov-Takens point, $GH$ is the generalized point of the Andronov-Hopf bifurcation.}
\end{figure}

Examples of the phase portraits at characteristic points labeled with numbers in Fig.3 are given in Fig.2. The simplest mechanism of the complete synchronization destruction is concerned with the saddle-node bifurcation of equilibrium states when the stable equilibrium and the saddle merge and an invariant curve responsible for the quasi-periodic oscillations appears (see the transition from point 1 to point 4 in Figs.2,3). Unlike the case of dissipative coupling \cite{b6}, the second pair of equilibrium points continues to exist. Decreasing of the signal amplitude leads to their mergence but on the lower branch of the saddle-node bifurcation line  $SN_{2}$.

The second mechanism becomes apparent if we move from point 2 to point 4 through point 3. It consists of the following. Initially, an unstable limit cycle appears as a result of the nonlocal bifurcation. This limit cycle converges to a stable equilibrium point which disappears through the inverse Andronov-Hopf bifurcation  $H_{2}$.

Let us consider a more complex situation when the locking of autonomous oscillators is observed (Fig.3b). In this situation, another mechanism for the destruction of synchronization is possible (see the transition from point 5 to points 6 and 7). Amplitude decreasing of an external signal leads now to the supercritical Andronov-Hopf bifurcation on the line $H_{1}$. As a result, stable focus becomes unstable and gives birth to stable limit cycle in the system. Furthermore, unstable limit cycle is born from a separatrix loop. (In fragment 6 this cycle is almost touching the separatrix.) Stable and unstable limit cycles come closer to each other and then merge and disappear on the saddle-node bifurcation line of limit cycles $SNC$. Remaining in the phase space unstable focus is transformed into an unstable node, merges with saddle and disappears on the line $SN$.

Note that for the two-parameter analysis the number of variants is sufficiently high and depends on the path in the parameter plane. At the same time, possible scenarios depend on the location of codimension-two points, the Bogdanov-Takens points $BT$, and generalized points of the Andronov-Hopf bifurcation $GH$, in which a supercritical Andronov-Hopf bifurcation turns into a subcritical one. The saddle-node bifurcation line of limit cycles $SNC$ comes also into the point $GH$.

In the considered situation, there are also new variants of the phase multistability. In the point 8, the stable invariant curve coexists with the stable equilibrium. Equilibrium state corresponds to the phase shift between oscillators equal approximately to $\pi$. In this case, out-phase oscillations of the oscillators are locked by an external signal. Stable quasi-periodic oscillations occur when the phases are approximately identical. Beyond the synchronization tongue (point 9 in Fig.2,3), there is a situation of coexistence of two stable invariant curves corresponding to the in-phase and out-phase locking of relative phase of oscillators. In similar way, in-phase oscillations of the oscillators are locked by an external signal near the first tip in Fig.3b.

Therefore, the bifurcation picture of a system with the reactive coupling differs significantly from that of a system with the dissipative coupling and is more complicated.

\section{Different types of regimes in the parameter plane}

Now we consider a more complete picture of different regimes possible in the system (\ref{eq10}). For this, we supplement the results of bifurcation analysis by the method of the charts of Lyapunov exponents. This method is highly effective for analysis of the systems with multi-frequency dynamics \cite{b5}. We calculate both Lyapunov exponents $\Lambda_{1}$ and $\Lambda_{2}$  of the system (\ref{eq10}) at each grid point on the parameter plane. Then we color these points in accordance with the values of Lyapunov exponents, so the following regimes are visualized:
\begin{enumerate}
  \item $P$ is a periodic regime with $\Lambda_{1}<0$ and $\Lambda_{2}<0$ (red color);
  \item $T_{2}$ is a two-frequency quasi-periodic regime  with  $\Lambda_{1}=0$ and $\Lambda_{2}<0$  (yellow color);
  \item $T_{3}$ is a three-frequency quasi-periodic regime with $\Lambda_{1}=0$ and $\Lambda_{2}=0$ (blue color).
\end{enumerate}

Obtained charts are given in Fig.4.  Fig.4a shows the situation when oscillators demonstrate beating regime in the absence of an external signal. In this case one can see the system of two-frequency synchronization tongues with the tips on the $\Omega$ -axis. These tongues are located inside the region of three-frequency quasi-periodicity $T_{3}$. Region of the periodic regimes $P$ has the threshold by  amplitude of the signal.
\begin{figure}[!ht]
\centerline{
\includegraphics[scale=0.32]{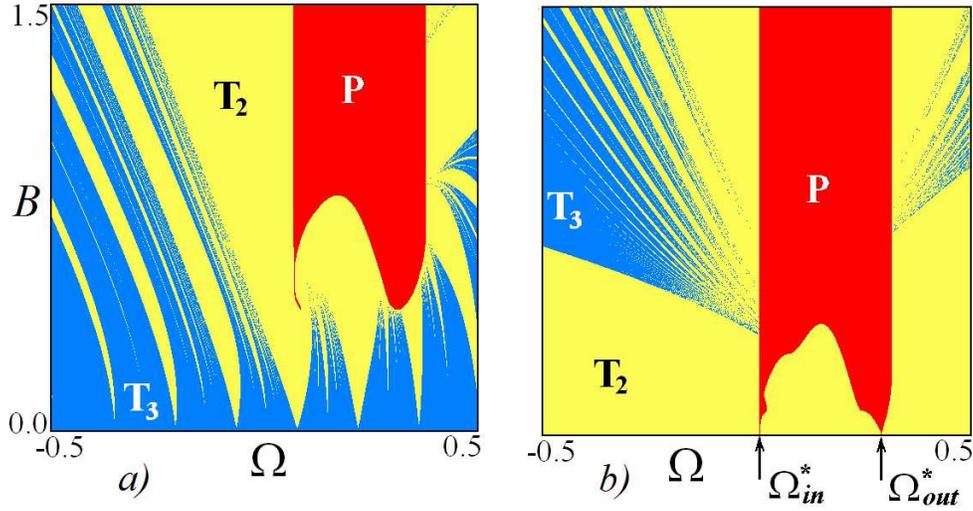}}
\caption{Charts of Lyapunov exponents for the phase system (\ref{eq10}); $\varepsilon = 0.3$;   a) $\Delta = 0.3$; b) $\Delta = 0.05$.}
\end{figure}

Another situation, when autonomous oscillators are locked in the absence of an external signal, is presented in Fig.4b. Here, the  region of the three-frequency regimes $T_{3}$ has the threshold by  amplitude of the signal. Inside this region, there is a set of narrow two-frequency resonance tongues of different type. The region of the periodic regimes $P$ touches the  $\Omega$-axis in two points. These points correspond to the resonances of an external signal with in-phase and out-phase modes, see Fig.1b. For small amplitudes of the external signal, only  two-frequency quasi-periodicity  is observed. Note that the chart of Lyapunov exponents reveals fine complex structure of the two tips in Fig.4b which is due to the fact that the region of complete synchronization is bounded by the saddle-node and Andronov-Hopf bifurcation lines.

\section{Dynamics of the initial system}

Let us give some illustrations for the initial system (\ref{eq1}). Its properties depend on the control parameter  $\lambda$. We set  $\lambda = 0.1$ which is much less than one. Therefore, one can expect a good agreement with the results for the phase model. On the other hand, this value is not very small, and possible differences may be observed. Corresponding chart of Lyapunov exponents is presented in Fig.5. Besides the regimes which are typical for the phase model, there are colored black chaotic regimes $C$ with positive largest Lyapunov exponent.
\begin{figure}[!ht]
\centerline{
\includegraphics[scale=0.32]{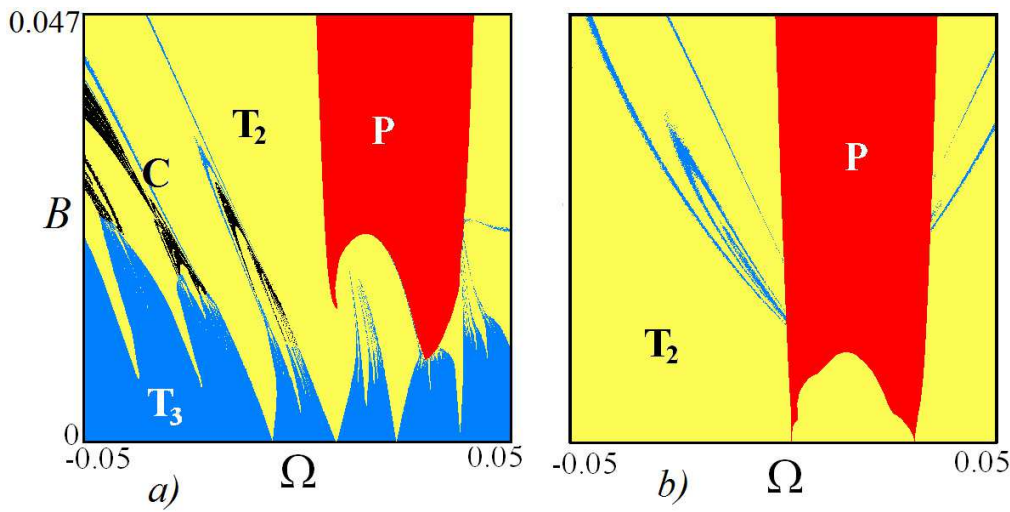}\\}
\caption{Charts of Lyapunov exponents for the system (\ref{eq1}); $\lambda = 0.1$,  $\varepsilon = 0.03$;   a) $\Delta = 0.03$; b) $\Delta = 0.005$.  If compare to Fig.4, all the parameters are normalized by  $\lambda$.}
\end{figure}

Comparison of Fig.4 and Fig.5 allows to conclude that the phase model describes the main elements of synchronization picture sufficiently well even at  $\lambda = 0.1$. Fig.4a and Fig.5a demonstrate not only the same configuration of the region of complete synchronization $P$ but also identical structure of the two-frequency tongues. At the same time, the tips of the region of complete synchronization looks similar in detail in Fig.4b and Fig.5b constructed for the case when autonomous oscillators are locked. Certain differences occur with increase of the signal amplitude $B$. (This is obvious for assumptions made for the phase model.) Thus, chaos occurs in the overlapping area of two-frequency tori  (Fig.5a). The region of three-frequency  regimes are replaced significantly by the region of two-frequency regimes in (Fig.5b). Note that the phase model will work better than less control parameter $\lambda$ and coupling constant $\varepsilon$.

Fig.6 shows projections of the attractors on the variable planes for the first and second oscillators of the system (\ref{eq1}). There is also ($x$, $y$)  plane which is suitable for visual type definition of the regime. Fig.6a represents the two-frequency torus, Fig.6b shows the three-frequency torus, and Fig.6c illustrates the chaotic regime. One can see that the regimes are sequentially complicated. For the invariant tori, attractor portraits look like the orbits similar to circles. This indicates the phase approximation efficiency. For the chaotic regime, orbit of the first oscillator is strongly disturbed, and trajectory  attends neighborhood of the origin. Now the phase model is not useful.
\begin{figure}[!ht]
\centerline{
\includegraphics[scale=0.5]{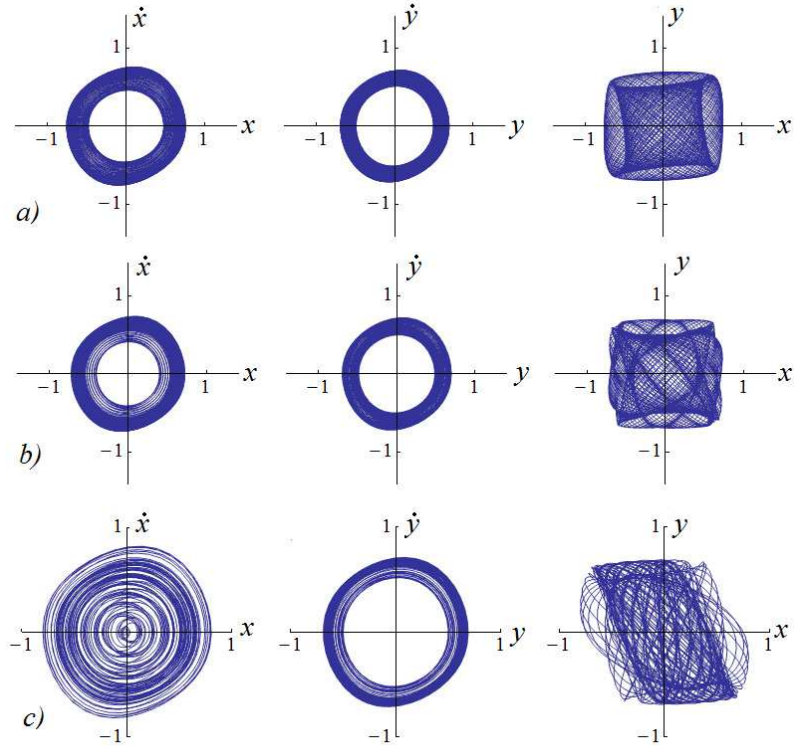}\\}
\caption{Examples of attractor projections for the system of three excited oscillators (\ref{eq1}); $\lambda = 0.1$, $\Delta = 0.03$, $\varepsilon = 0.03$. a) Two-frequency torus, $\Omega = 0.975$, $B = 0.04$; b) three-frequency torus, $\Omega = 0.95$, $B = 0.01$,   c) chaotic attractor, $\Omega = 0.975$, $B = 0.0375$.}
\end{figure}

Note that the phase model will work better than smaller coupling constant $\varepsilon$. We will reduce tripled the coupling constant that would confirm this. The corresponded chart of Lyapunov exponents is presented in Fig.7.  From the comparison of Figs. 7 and 4b, we see that in this case the phase model becomes practically precision.
\begin{figure}[!ht]
\centerline{
\includegraphics[scale=0.35]{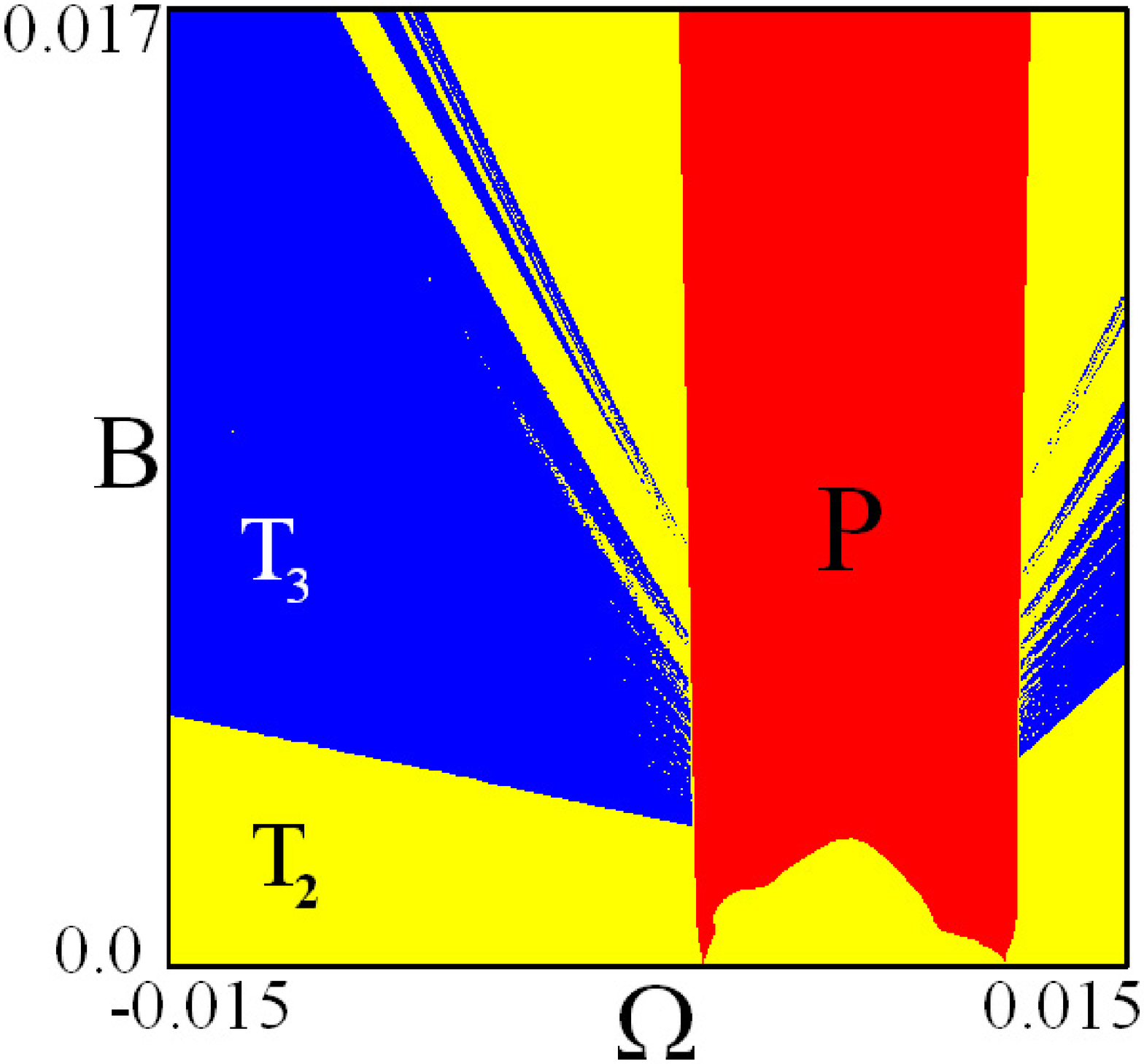}\\}
\caption{Charts of Lyapunov exponents for the system (\ref{eq1}). $\lambda = 0.1$,  $\varepsilon = 0.01$ and $\Delta = 0.000556$.}
\end{figure}

\section{Conclusion}

When describing synchronization of two reactively coupled oscillators by an external signal, we need to consider effects which are of the second order in coupling parameter. Locking of both oscillators by an external signal can be accurate or not. When the locking is accurate, there is a stable equilibrium in the phase equations. For the non-accurate locking, there is a limit cycle on the phase plane. An external signal of small amplitudes can lock both the in-phase and out-phase oscillations of the oscillators. Basic bifurcation mechanisms of the synchronization destruction are as follows: merging of the stable and unstable equilibriums; subcritical Andronov-Hopf bifurcation which results in disappearing of the stable equilibrium; saddle-node bifurcation of the stable and unstable limit cycles. The bifurcation picture includes also the codimension-two bifurcations: Bogdanov-Takens and generalized Andronov-Hopf bifurcation points. A feature of the system of forced reactively coupled oscillators is an ability of the phase bistability which consists in coexistence of different types of stable regimes (equilibrium points, limit cycles, invariant curves).

\textit{The authors acknowledge support of the RF President program for leading Russian research schools NSh-1726.2014.2. Yu.V.S. also thanks Russian Foundation for Basic Research (grant No.14-02-31064).}

\end{document}